# Deep neural networks for collaborative learning analytics: Evaluating team collaborations using student gaze point prediction

Zang Guo and Roghayeh Barmaki
Computer and Information Sciences, University of Delaware
Correspondence to: Roghayeh Barmaki, E-mail: rlb@udel.edu

Automatic assessment and evaluation of team performance during collaborative tasks is key to the learning analytics and computer-supported cooperative work research. There is a growing interest in the use of gaze-oriented cues for evaluating the collaboration and cooperativeness of teams. However, collecting gaze data using eye-trackers is not always feasible due to time and cost constraints. In this paper, we introduce an automated team assessment tool based on gaze points and joint visual attention (JVA) information extracted by computer vision solutions. We then evaluate team collaborations in an undergraduate anatomy learning activity (N=60, 30 teams) as a test user-study. The results indicate that higher JVA was positively associated with student learning outcomes ($r_{(30)} = 0.50, p < 0.005$). Moreover, teams who participated in two experimental groups, and used interactive 3-D anatomy models, had higher JVA ($F_{(1,28)} = 6.65, p < 0.05$) and better knowledge retention ($F_{(1,28)} = 7.56, p < 0.05$) than those in the control group. Also, no significant difference was observed based on JVA for different gender compositions of teams. The findings from this work offer implications in learning sciences and collaborative computing by providing a novel mutual attention-based measure to objectively evaluate team collaboration dynamics.

*Implications for practice and/or policy:*
- Joint visual attention skills are most crucial for children with attention deficits, specifically those in the Autism spectrum. Our joint attention assessment method can be applied to both social and non-social interactions present in video data.
- Psychologists, learning scientists, and experimental designers can use our method to objectively evaluate the quality of learning or collaborative processes without constraints of one-time, self-reported assessments.

*Keywords:* collaborative learning analytics, tertiary education, co-located team collaboration, gaze following, joint visual attention, deep learning.

## Introduction

Collaborative learning is an essential educational instrument for teaching and learning. As a team-based and student-centered educational practice, it promotes student motivation and enhances knowledge retention via teamwork and cooperation (Sung & Hwang, 2013). While collaborative learning has been widely introduced and practiced in various situations in recent years, measuring and evaluating collaboration remains a challenge. Fairness, rationality, and automatism are the core issues that need to be considered during the analytics (Bertsimas & Gupta, 2015).

Gaze-oriented cues can be used as a means of obtaining information about the cognitive activities of a collaborator, and there is evidence that students look at and point to the same object for collaborations during learning activities (Schneider & Pea, 2013; Schneider et al., 2018). This gaze alignment is called Joint Visual Attention or JVA (Van Rheden et al., 2017)—see Figure 1, for example. JVA is noted as a strong predictor of successful collaboration among students (Pietinen et al., 2010; van der Meulen et al., 2016). Compared with collecting traditional self-reported survey data or collaboration system data for the one-time performance evaluation (Wendler, 2006), capturing gaze alignment during the entire collaborative process with the JVA criterion can reveal more valuable information about the quality of interactions among teams (Wahn et al., 2016). New technologies provide innovative methods to extract and measure JVA features. A growing number of researchers take advantage of sensor-based eye-trackers to objectively measure gaze features during various social interactions, especially for co-located collaborative tasks

(Huang et al., 2019; Trail, 2019). However, despite the provision of highly accurate data from eye-tracking devices, these sensors are usually highly-priced and may introduce limitations for educational study settings, specifically those needed to be carried out in classrooms and not in research labs. For example, the calibration might be a time-consuming process, or the study activity is needed to be completed within the limited tracking range of the sensors.

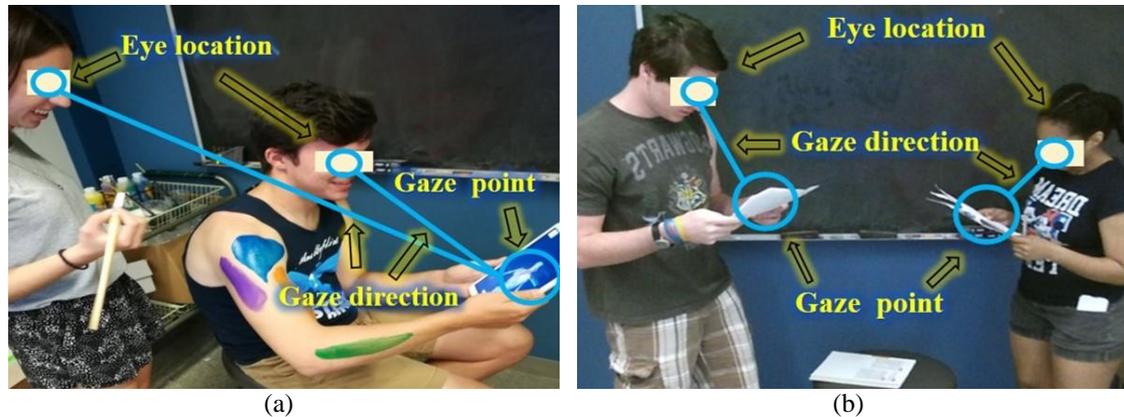

*Figure 1.* Examples of the gaze following method in our study (a) with JVA feature: students' gaze points converge on the tablet, and (b) without JVA feature: students are looking at their own notes.

With the emergence of different deep learning techniques, neural networks can approach the gaze tracking problem differently. Using a deep learning framework to track the gaze features from a sequence of 2-D images or videos is practical and robust in understanding and interpreting students' behaviors in human-human interaction and human-object interaction (Recasens et al., 2015). For example, when two students are looking for a path from the library to the gym on a single university map, we can easily find out if they are sharing the same information, or predict whether they will pick the same path by following their gaze direction. Compared with eye tracking and gaze estimation, the gaze following method (Hansen & Ji, 2009; Santini et al., 2017; Lian et al., 2018) not only estimates the gaze direction but also predicts the gaze point from the image without the need for specialized hardware (e.g., head-mounted camera, infrared light source) and tedious gaze calibration procedure. Figure 1 shows the application of gaze following method on top of two images captured from our test user-study.

In this paper, we introduced a computer vision-based solution for team performance evaluation using mutual gaze point predictions, along with a collaborative anatomy learning activity as a test user-study for our approach. Collaborative activity sessions were recorded by a color camera as a sequence of images. For collaboration analysis, we first tracked team members' gaze directions, and the focus objects during the activity using the gaze following method (Lian et al., 2018). We then extracted JVA features of teams and analyzed it with other collected data, including post-test scores and demographics information related to team gender compositions. This study hypothesizes that students who share mutual gaze during the activity—e.g., those teams with higher JVA values—obtain higher scores in their post-activity knowledge tests as well since they engage more in collaborative tasks. We are also interested in understanding if JVA values are varied significantly in (a) different study conditions, and (b) different gender compositions of teams for collaborative learning.

This paper is organized as follows: First, we review the literature on the related work of anatomy education, collaborative learning analytics, JVA applications, and gaze following approaches. We then introduce details of gaze following method and our proposed assessment measures for collaboration. Finally, we both present findings from our test user-study and discuss further implications.

## Literature review

### Educational technology for anatomy education

In the domain of human anatomy learning, different education technologies have recently replaced traditional teaching methods such as lectures, cadavers, and textbooks. With modern computer-assisted technologies, 3-D visualization methods improve students' performance by allowing them to explore 3-D anatomical models on 2-D mobile device screens (Yammine & Violato, 2015). Mobile-based applications and web-based 3-D games have been used as learning tools for the study of human skeletal, muscular, and cardiovascular systems to name a few (Golenhofen et al., 2019; Lemos et al., 2019). Virtual Reality (VR) and Augmented Reality (AR) techniques have been adopted into medical education and surgical training fields in recent years (Nicholson et al., 2006; Silva et al., 2018; Maresky et al., 2019). As powerful learning tools, VR and AR engage students in an immersive environment with audio and visual interactions, and stereoscopic 3-D models to enhance their learning experience (Luursema et al., 2006, 2008; Hackett & Proctor, 2018). In this paper, we have evaluated team performance in a controlled study that leverages modern anatomical content visualization in 3-D with handheld mobile/tablet devices and large-scale AR displays.

**Gender effects and collaboration**

As the results of the different socialization processes, gender differences have been discussed by recent studies on the team level. The relationship between gender and collaboration is not uniform, and it varies based on different disciplines and tasks (Wegge et al., 2008; Fernandez-Sanz & Misra, 2012). Previous researches have shown that females have better information-processing skills than males during cognitive tests (Schaie & Willis, 1993; Rabbitt et al., 1995). Females' higher management ability in collaborative tasks was also highlighted (Eagly & Carli, 2003; Bear & Woolley, 2011; De Paola et al., 2018). Several studies concluded that collaboration performance would be improved with females involved. Other research showed that women often had negative experiences on teams due to gender biases at the technical level, especially in science, technology, engineering, and mathematics (STEM) (Meiksins et al., 2013; Meadows et al., 2015). Conversely, some reported no significant gender effect. Andersson (2001) argued that, although females have better performance on individual memory tasks, no main gender effect was found in collaborative tasks. Prinsen et al. (2007) noted that females were more likely to collaborate, and males were more assertive in computer-mediated communication (CMS, (Herring, 1996)), and computer-supported collaborative learning (CSCL, (Lipponen, 1999)) settings. In the same study, Prinsen et al. also acknowledged that different distributions of roles in collaborative learning might change gender contributions. We will explore potential gender differences in our anatomy learning study in association with learning outcomes and joint attention measures.

**Collaborative learning analytics**

The importance of social interactions during the learning process has been emphasized in the past (Okita et al., 2007). Collaborative learning not only helps students to improve teamwork skills, but also promotes learning motivation, increases learning experience, enhances brainstorming skills (Webb et al., 1995), and facilitates their learning performance during team interactions (Sung & Hwang, 2013). Consequently, researchers highlighted that instead of using new learning formats, one should pay more attention to the measurements of and access to collaboration performance (Huang et al., 2019). In early attempts to analyze collaborative learning, Soller and Lesgold (1999) provided a practical collaborative learning framework and evaluated active learning skills using conversational interaction data collected from surveys. With the proposal of CSCL research, several machine learning techniques were used to predict student grades using support vector machines (Baker & Yacef, 2009), decision trees (Rosé et al., 2008), and regression (Kotsiantis, 2012) to name a few. Those solutions either established effective collaborative learning models or built reasonable standards for judging collaboration performance based on single-time solicitation techniques. However, the data used in those models were collected from class attendance, quiz scores, or reports, which may only represent students' one-time or episodic performance during the learning activities.

For image and video data analysis of teamwork quality assessment, an object detection approach was introduced in previous work (author, 2019) to extract useful features such as the locations of students and objects in the scene. That study provided a new perspective, using a deep learning method to process image data sequences to analyze co-located teamwork.

**Joint visual attention applications**

Joint visual attention features have been lately introduced to a range of applications including collaborative search (Brennan et al., 2008), mediated interaction (Bente et al., 2007), infant-caregiver interaction (Markus et al., 2000), and training for children with autism (Whalen & Schreibman, 2003). Recently, it has been a growing interest on the use of synchronized eye-trackers to quantitatively measure gaze alignment in various collaborative situations of interpersonal communication (Van Rheden et al., 2017; Bryant et al., 2019; Huang et al., 2019; Kim et al., 2020). However, there are main challenges of using eye-tracking sensors, including the high cost of the devices, and restricted environmental and calibration settings (e.g., the camera should be precisely in front of the student within a close distance, and on top of a specific panel (Huang et al., 2019)). Image-based computer vision methods—as a more affordable alternative approach— have also been used for extracting gaze features in recent studies. Yücel et al. (2013) presented an image-based head pose estimation method for establishing joint attention between an experimenter and a robot using a color camera. Later, Harari et al. (2018) identified the common gaze target by combining the estimated 3-D gaze direction with image segmentation.

### Gaze following using deep learning

There has been an expanding interest in the estimation and reconstruction of human gaze direction from 2-D images to identify their activities in the scene using various deep learning frameworks. Gaze following is the task of following people's gaze in a scene and inferring what they are looking at. Compared with eye tracking and gaze estimation, gaze following not only estimates the gaze direction but also detects the focus point from the image (Lian et al., 2018). Patacchiola and Cangelosi (2017) proposed a face detector to extract face landmarks and estimate head poses using convolutional neural networks. Marín-Jiménez et al. (2014) used head pose detection with implicit pose information to detect human-human interaction in videos. However, those previous works were limited by the complexity of inputs (massive eye-tracking data (Yücel et al., 2013)); restricted situations (resolution of the image (Marín-Jiménez et al., 2014)); and field of view (the distance between the camera and students (Zhu & Ramanan, 2012)). In the work of Recasens et al. (2015), the gaze point of multiple observers in daily scenarios was predicted using deep neural networks and saliency models of attention. Mukherjee and Robertson (2015) combined RGB-D images and multimodal data to reconstruct 3-D head poses and follow gaze direction in images and videos. These studies motivated the work reported here, using a deep learning approach to target gaze alignment features for the novel application of collaborative learning analytics.

We are interested in understanding how two students are interacting with one another, or with objects, and following the gazes of multiple observers in a scene. Hence, we use gaze-following methods (Lian et al., 2018) to estimate both the gaze direction and the gaze points to collect human-human interaction information. Further details about our approach are presented in the following sections.

## Method and materials

### Intervention

We conducted a between-subjects study of collaborative muscle learning intervention in a laboratory course of General Biology as part of undergraduate premedical curricula. Over 300 students in 134 teams participated in the original study, and the data from a subset of teams with size two (N=60, 30 teams) was selected as our test dataset. Students worked in teams to complete a muscle painting activity as part of their required lab activities. During this intervention, students worked together to learn more about human muscle groups in a body painting activity. They were expected to identify and paint major muscles of their body using one of the learning instruments (textbook or lab manual, tablet, and AR) and washable painting supplies. The first student played the role of a model, while their teammate, as a painter, located the major upper-limb muscles with the aid of their lab manual (Marieb & Jackson, 2006) or other digital devices, and painted the model's upper limb. Afterward, students switched roles, and the upper limb painter became a model for the lower limb. The goal of this learning activity was to make sure all students could gain knowledge of anatomy in a collaborative effort. See Figure 2 to learn more about the intervention details.

As briefly mentioned, our study had three different settings based on instrumental tools. Students in the control group used textbooks as their learning tools. In experimental group I, instead of a textbook, students used our in-house interactive app on the tablet as a 3-D musculoskeletal visualizing system. Experimental group II used a screen-based AR system—also developed internally—where students could see themselves with augmented anatomy visualizations on a large display. The knowledgebase information presented in all instrumental tools were identical across the board to mitigate potential confounding factors related to student workload and learning. There was also a mobile workstation with a Kinect sensor inside the laboratory room to capture snapshots from students during learning activities. Figure 2 shows our three study conditions of the learning activity.

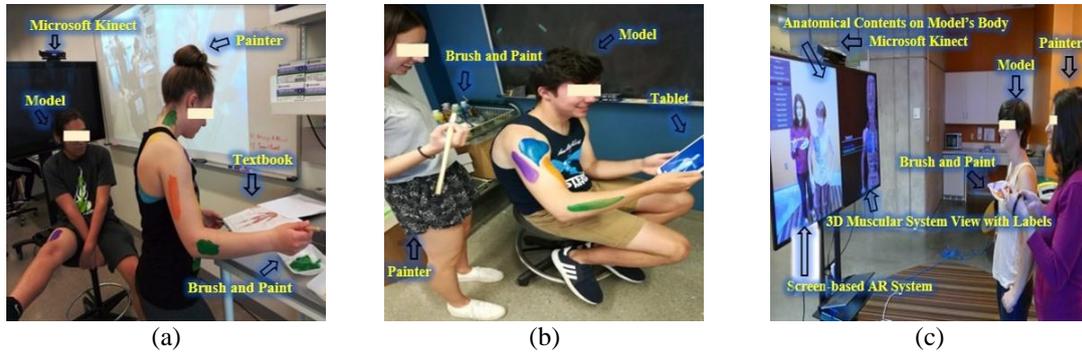

*Figure 2.* Study conditions for students in pairs to complete anatomy painting intervention using a/an (a) textbook, (b) interactive app on the tablet, or (c) screen-based AR system.

The study was approved by the Institutional Review Board, and oral informed consent was obtained from each participant student before the study commenced. After consent, students entered the activity room with their teammates and completed the task. All students completed both pre- and post-activity questionnaires and knowledge tests.

**Datasets**

*Surveys.* Using the Qualtrics application, survey data was collected from all students individually after completing the activity. This survey included demographics information, usability questions, and a post-test about the human muscle system.

*Image training data.* The large-scale gaze-following dataset, the GazeFollow dataset, used for training came from Recasens et al. study (2015). This benchmark dataset included 130,339 people and 122,143 images in total with gaze points inside the image.

*Image test data.* The test dataset included a total of 4,646 images collected from 30 pairs of students during the collaborative learning activity in three conditions (ten teams from each condition of the textbook, tablet, and AR, totaling images from 30 teams). Images were captured every ten seconds, and each image file was timestamped. The resolution of each test image was 2560 × 1440 pixels. Images with camera difficulties or additional individuals in the scene were discarded.

**Gaze following framework**

To extract shared gaze features from the images, we needed to estimate the students' gaze direction and focus in the scene. Thus, we applied a two-stage gaze following approach (Lian et al., 2018) on our test dataset. This method was very suitable for our project since it was capable of detecting the gaze direction from the head image and predict the potential gaze point along the gaze direction, via a deep learning framework. The gaze following approach and its underlying network architecture is shown in Figure 3.

The idea was inspired by human behaviors of gaze following (Lian et al., 2018). First, a gaze direction was estimated from the gaze direction pathway. In the gaze direction pathway, the resized head image (224 × 224)—image sizes are listed in pixels hereafter—was fed into the convolutional neural network ResNet-50 (Rezende et al., 2017) for feature extraction. Then the head features were concatenated with head position features encoded by one fully connected layer for gaze direction estimation. A coarse gaze direction was predicted as the vector output and then encoded as multi-scale gaze direction fields. The gaze point was supposed to be in the gaze direction. Next, the multi-scale gaze direction fields were combined with scene contents (224 × 224) and fed into the heatmap pathway for heatmap regression using a Feature Pyramid Network (Zhao et al., 2019). The heatmap (56 × 56) represented the probability distribution of the gaze point, and the point with the maximum value of the heatmap represented the probable gaze point of the scene.

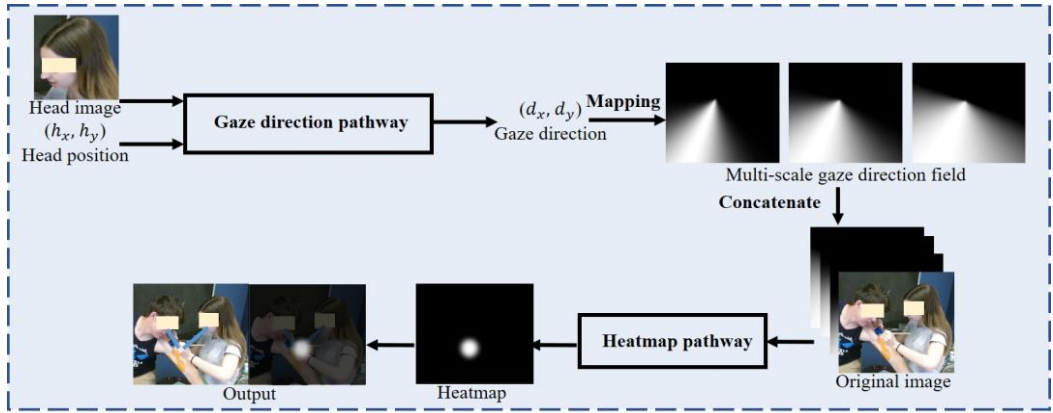

*Figure 3.* The network architecture for the gaze following method (Lian et al., 2018) atop our collaborative study image frames. Using the heatmap we can predict the gaze point convergence of students in the collaborative activity.

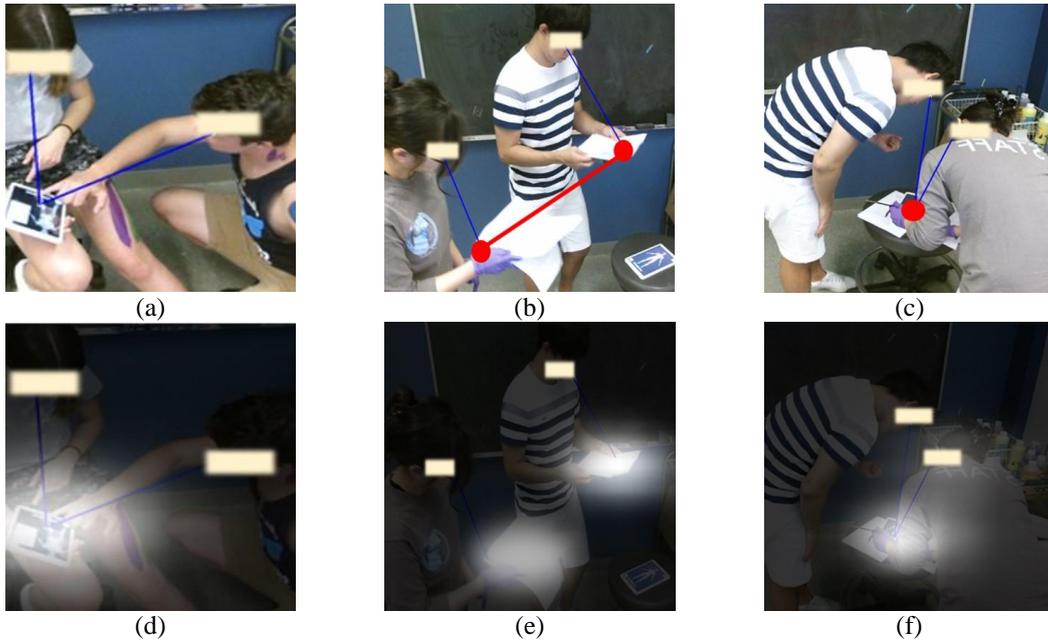

*Figure 4.* Gaze following results for three sample frames: (a) gaze following output to show gaze directions of individuals with blue lines; (b) output without JVA feature (Euclidean distance between the gaze points of students is greater than 100 pixels; (c) output with JVA feature (Euclidean distance between the gaze points is smaller than 100 pixels); and (d-f) heatmaps associated with the gaze points. While two distinct sheets of papers are predicted as gaze points for team members in (e), (d) and (f) are examples of the convergence of visual attention on the tablet device, thus JVA feature is true.

According to the authors (Lian et al., 2018), this gaze following approach outperformed other existing methods on gaze point prediction. Compared with state of the art, this method decreased 23.68% of the Euclidean distance error for gaze point on the GazeFollow dataset. We chose gaze following approach in our method because it managed to simulate the third view person's gaze following behavior, and it was robustly trained by using the heatmaps for focus points to assist gaze direction estimation.

The gaze following output shown in Figure 4.a visually draws a blue gaze line on the original image for each individual in the scene. The blue line is initiated at the eye location and terminated at the predicted final gaze point. The highlighted region in the corresponding heatmap (Figure 4.d-f) represents the predicted gaze point where the students are looking at. The output also marks the coordinates of each gaze point, which is used in our approach as a collaboration metric. We are interested in the automatic recognition of joint or mutual gaze visual attention among students in every image sequence during the collaborative task. Further details about JVA feature analysis as a collaboration measure are presented in the following section.

**Evaluation measures**

We have analyzed team performance and collaboration based on objective measures related to joint attention, knowledge retention, study conditions, and gender composition of teams. These evaluation measures of collaboration are described as following.

*JVA ratio.* JVA represents the shared focus of two or more individuals and plays a key role in collaboration prediction (Bruinsma et al., 2004). In this work, we defined the JVA ratio for each team based on the frequency with which the two students shared gazes during the collaborative activity, divided by total image frames captured from the team—a normalized measure of JVA based on total activity frames of teams. Since there was lots of cooperation between painters and models during the learning activity process, they needed to share attention and maintain mutual gaze most of the time: while painting, discussing, and looking at the learning materials. For example, when the painter was painting, both painter and model may have looked at the same location, the active painting region. When students needed to find the muscle's correct location, the painter and the model may have shared the screen of the interactive app on the tablet or the AR view to zoom in the 3-D musculoskeletal system.

We used Euclidean distance between the gaze points detected by the gaze following method for automatic identification of JVA in each image frame (Is_JVA was a Boolean variable per each frame, was set to false by default). Based on image size and resolution, we recognized JVA, or mutual gaze feature and set Is_JVA as true, if that distance was smaller than 100 pixels. For each team, the JVA ratio was computed by the total number of frames in which the Is_JVA variable was true, divided by the total number of frames. Some examples with and without JVA recognition are shown in Figures 1 and 4. We were interested to learn if a higher JVA ratio is also associated with better learning outcomes.

*Team post-test score.* The key objective of this collaborative learning activity was to enhance anatomy knowledge retention of students. All participants needed to independently—not with the assistance of their peers—locate and label five muscle names in a diagram of the human musculature in the post-test; thus, individual test scores ranged between 0 and 5 with discrete values. Since we used the average of post-test scores per team and named it team post-test score, the team post-test score was still in the same range, but non-discrete values were also observed in the dataset.

*Study conditions.* As mentioned earlier, there were three different conditions or settings for our muscle painting study. Students in the control group used textbooks as their learning tools. The experiment groups either used a tablet or the screen-based AR system to complete the task. We wanted to investigate the differences in team performance based on these three conditions and two groups of experimental and control.

*Gender composition.* Students were pre-assigned randomly to teams to complete the muscle painting activity. There were three possible gender compositions in per pair of students: male pair, female pair, and mixed pair. We were interested in evaluating the gender effects in collaborative learning and investigating if any significant variability of JVA ratios and knowledge tests was present in female-female, male-male, and male-female (mixed) pairs of students.

# Results

In the following, we report descriptive and inferential results from our test user-study. We particularly look at the JVA ratio—an automatically generated measure based on our proposed framework using deep neural networks—in association with our evaluation measures. Table 1 summarizes descriptive statistics for pairs of students in each study condition, including the number of teams, mean values, and standard deviations for JVA ratios and team post-test scores.

*Participants.* We had analyzed data from 60 participants (38 Females) in 30 teams in this work. All of these students were enrolled in the undergraduate pre-medical program in the affiliated university. There were ten teams per the condition of the study—textbook, tablet, and AR. Knowing that tablet and AR conditions were part of the experimental group, we had 20 teams in the experimental group and 10 in the control group. Data from teams with a larger size, those with students under 18 years of age, and those with incomplete data were excluded in this study.

*JVA ratio.* JVA ratio was the percentage of the time teams had shared mutual or joint attention during the learning activity. Although no significant difference between the three study conditions and the JVA ratio was observed, the p-value was very close to the critical value of α ($F_{(2,27)} = 3.26, p = 0.054, ns$—$ns$ stands for statistically non-significant). Interestingly, the JVA ratio of two experimental groups of tablet and AR ($n = 20, M = 45.6, SD = 15.97$) was significantly higher than those in the control condition who used textbook ($n = 10, M = 31.3, SD = 9.73$), and this finding was statistically significant with a large effect size ($F_{(1,28)} = 6.65, p < 0.05, Cohen's\ d = 1.00$ (Cohen, 2013)). Table 1 and Figure 5 provide additional information about JVA ratio distribution across all study conditions and groups.

*Team post-test score.* A significant difference based on team post-test scores was observed among study conditions of textbook ($M = 1.15, SD = 0.95$), tablet ($M = 2.35, SD = 1.03$), and AR ($M = 2.35, SD = 1.20$), ($F_{(2,27)} = 3.64, p < 0.05, r^2 = 0.16$, medium effect size). Post-hoc comparisons indicated pairs of textbook and tablet, and textbook and AR conditions were different from each other based on differences of means. Similarly, the team's average post-test score from two experimental groups of tablet and AR ($M = 2.35, SD = 1.20$) was significantly higher than those in the control group ($M = 1.15, SD = 0.95$), and this finding was statistically significant with a large effect size ($F_{(1,28)} = 7.56, p < 0.05, Cohen's\ d = 1.06$).

Table 1.
*Summary of JVA ratio and team post-test score with different instrumental tools*

| Group | Condition (Instrumental tool) | Observation (teams) | JVA ratio (%) | Team post-test score |
|---|---|---|---|---|
| | | $n$ | M ± SD | M ± SD |
| Control | Textbook | 10 | 31.30 ± 9.73 | 1.15 ± 0.95 |
| | Tablet | 10 | 46.50 ± 15.43 | 2.35 ± 1.03 |
| | AR | 10 | 44.60 ± 17.28 | 2.35 ± 1.42 |
| Experiment | Combined (Tablet and AR) | 20 | 45.55 ± 15.97 | 2.35 ± 1.20 |
| **Total** | Textbook, Tablet, AR | 30 | 40.80 ± 15.59 | 1.95 ± 1.25 |

Note. JVA = Joint visual attention; M = Mean; n = Number of observations; SD = Standard deviation; Team post-test score is in range from 0 to 5; JVA Ratio (%) is in range from 0 to 100.

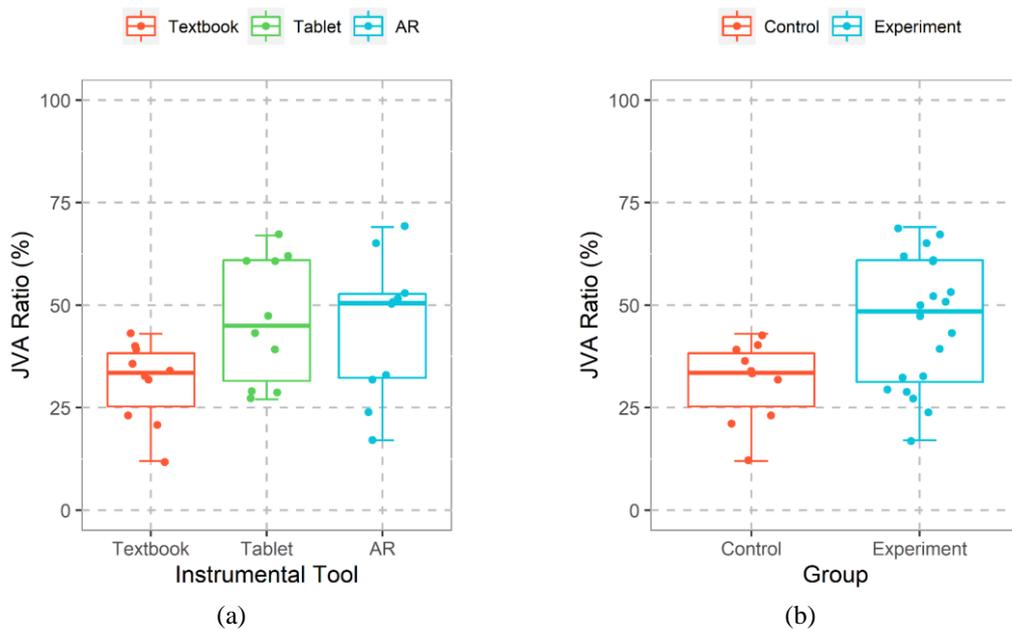

*Figure 5.* The boxplot with observed data points for JVA ratio across (a) different study conditions/instrumental tools of textbook, tablet, and AR, (b) different groups of control and experiment. JVA ratio was significantly different for control and experimental groups.

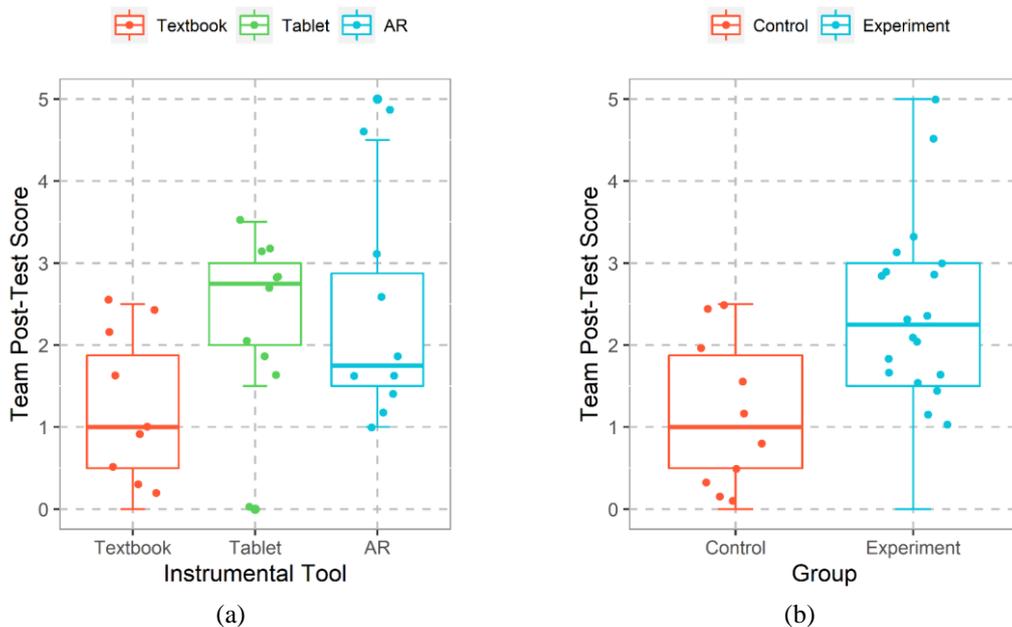

*Figure 6.* The boxplot with observed data points for team post-test score across (a) different study conditions/instrumental tools of textbook, tablet, and AR, (b) different groups of control and experiment. A significant difference was observed both in the study conditions and groups for post-test scores.

*JVA ratio and team post-test score.* We also measured the association between JVA and team post-test scores using the Pearson correlation coefficient (Benesty et al., 2009). The Pearson correlation measure indicates a significant positive linear association (Zou et al., 2003) with a strong relationship between the JVA ratio and team post-test scores ($r_{(30)} = 0.50, F_{(1,28)} = 9.33, p < 0.005, r^2 = 0.25$ (large effect size)). The scatter plot drawn from data is shown in Figure 7. This finding shows that JVA features are strongly associated with learning outcomes, such as post scores. Points on the scatter plot closely resemble a straight line with a positive slope, which shows that post-test scores increase with higher JVA ratios. Therefore, the team with a high frequency of sharing gaze is more likely to achieve better outcomes in the post-test.

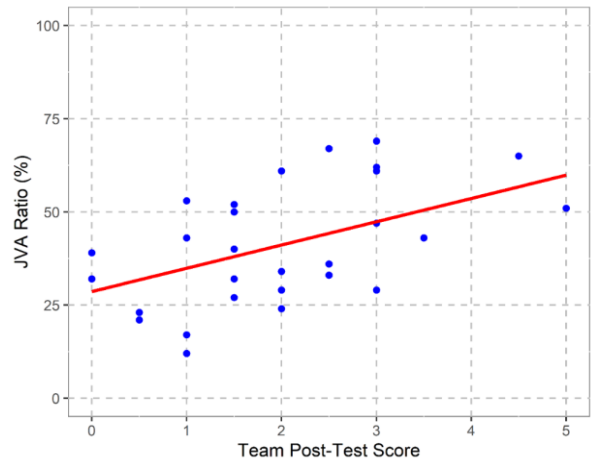

*Figure 7.* The scatter plot of JVA ratio with team post-test scores. The Pearson correlation and its underlying regression model indicate a significant positive correlation between JVA ratio and team post-test score.

*Gender composition.* Among 30 teams, we recorded the gender composition for each team from the survey data and investigated the gender effects on collaborative learning during the activity (see Table 2 and Figure 8 to learn more). Overall, mixed pairs (8 teams) achieved the highest JVA ratio ($M = 47.0, SD = 15.46$) and the best learning outcomes from post-test scores ($M = 2.50, SD = 1.28$), but this variability was not statistically significant ($F_{(2,27)} = 1.29, p = 0.29, ns$). Moreover, no significant difference was observed based on gender composition in teams for JVA ratios ($F_{(2,27)} = 1.10, p = 0.35, ns$).

Table 2.
*Summary of JVA ratio and team post-test score with different gender conditions*

| Gender composition | Observation (teams) | JVA ratio (%) | Team post-test score |
|---|---|---|---|
| | $n$ | $M \pm SD$ | $M \pm SD$ |
| Females | 15 | $37.00 \pm 15.05$ | $1.63 \pm 1.29$ |
| Males | 7 | $41.86 \pm 16.72$ | $2.00 \pm 1.04$ |
| Mixed | 8 | $47.00 \pm 15.46$ | $2.50 \pm 1.28$ |
| **Total** | 30 | $40.80 \pm 15.59$ | $1.95 \pm 1.25$ |

Note. JVA = Joint visual attention; M = Mean; n = Number of teams; SD = Standard deviation; Team post-test score is in range from 0 to 5; JVA Ratio (%) is in range from 0 to 100.

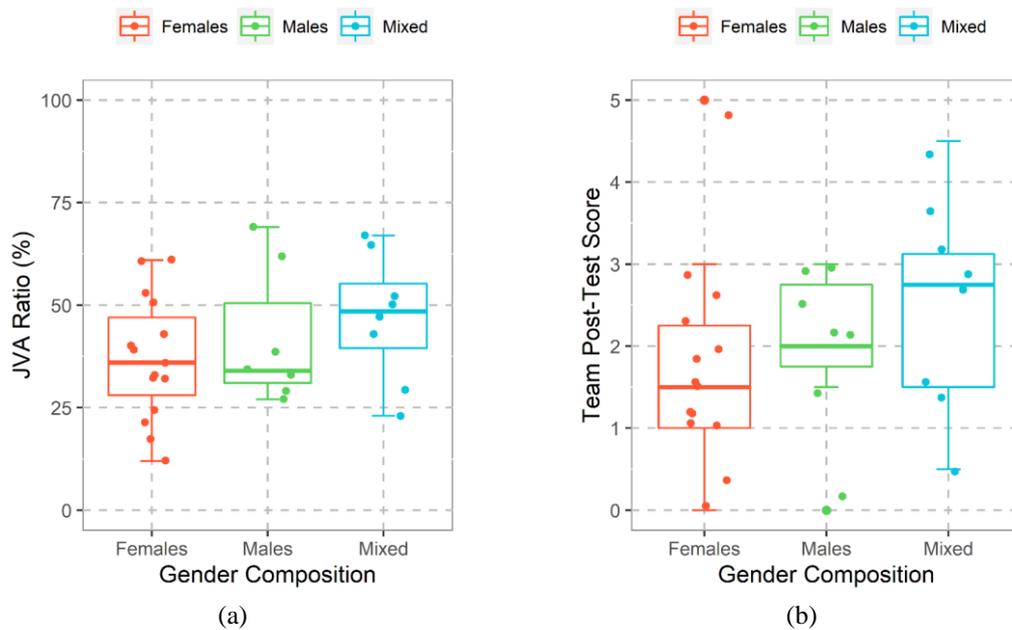

*Figure 8.* The boxplot with observed data points across teams with different gender compositions: (a) JVA ratios, (b) team post-test scores. No significant difference was observed in the study for JVA ratios nor post-test scores for different pairs of students. Among these 30 pairs or teams of participants, there were 15 female-female, seven male-male, and eight mixed pairs in total.

## Discussion

*JVA ratio*. Capturing gaze alignment during the collaborative process with JVA criterion can reveal valuable information about the quality of interactions among teams (Markus et al., 2000; Bruinsma et al., 2004; van der Meulen et al., 2016; Wahn et al., 2016; Bryant et al., 2019); however, not many studies have investigated computer-vision based approaches to better measure and capture it in co-located team-based learning interactions. In this study, we introduced a novel assessment tool for automatic team performance evaluation using mutual gaze information using the gaze following method (Lian et al., 2018). Compared with other methods using traditional one-time performance evaluation (Wendler, 2006) or high-cost eye-tracking devices (Bryant et al., 2019), our method was able to automatically extract JVA features during the whole learning process with a simple color camera. We also investigated the effectiveness of our JVA method in a test user-study. Results showed that JVA ratios of two experimental groups of tablet and AR were significantly higher than those in the control group who used the textbook. The findings are supported by previous research that looked at e-textbooks as a potential alternative learning tool based on gaze information from student users (Gelderblom et al., 2019), although this research was limited to individual learners and not teams.

*Team post-test score*. Post-test scores indicate that how much students achieved from the learning activity (Dimitrov & Rumrill Jr, 2003). In this study, we set up three different study conditions by using different instrumental tools for anatomy learning activity. Right after the activity, post-test scores were collected from students using a survey completed individually, and team post-test score was calculated as the average of team members' individual test scores. Team post-test scores of two experimental groups of tablet and AR were significantly higher than those in the control condition who used the textbook. This finding is in agreement with previous studies in anatomy education which highlighted the potential of using evolving technologies such as mixed and augmented reality for enhancement of student learning and student outcomes in anatomical science education (Nicholson et al., 2006; Silva et al., 2018; Maresky et al., 2019).

*JVA ratio and team post-test score*. There was a significant positive linear association with a strong relationship between the JVA ratio and the team post-test score. This is in agreement with our hypothesis of this study: students who shared mutual gaze with their teammates for a longer time on the learning task were more likely to obtain higher scores in their post-activity knowledge tests. This finding agrees with previous researches on the positive effects of sharing a gaze with learning outcomes, including imitation and socio-cognitive performance (Carpenter & Tomasello, 1995; Hirotani et al., 2009).

*Gender composition.* There were three possible gender compositions per pair of students: females, males, and mixed pairs. About half of the participants were in females-only teams, which was also a common gender enrollment rate in life sciences and premedical programs (Barzansky, 1997; Brooks, 2017). Even though mixed teams had slightly higher JVA ratios and better learning outcomes as shown in the results, no significant difference was observed based on the gender composition of teams neither for the JVA ratio, nor the team post-test score during the activity. These findings are consistent with previous studies that noted no gender effect in life sciences studies (Andersson, 2001; Prinsen et al., 2007).

In addition, we found that compared with control groups using text and 2-D anatomy models from the textbook, the students in both experimental groups had higher JVA ratios and better knowledge retention by interacting with 3-D models on the tablet screen or AR system. Specifically, in teams with the screen-based AR, students could easily collaborate and locate specific muscles with high accuracy projected on top of their own bodies. This outcome was also highlighted in a recent meta-analysis study as collaborative learning being the most beneficial approach in any AR interventions (Garzón et al., 2020). Our study also provides further evidence that 3-D visualization technologies increase students' engagement and improve their knowledge retention in human anatomy learning (Nicholson et al., 2006; Luursema et al., 2008; Yammine & Violato, 2015; Hackett & Proctor, 2018).

## Conclusion

In this paper, we introduced a novel computer vision-based measure for automatic team performance evaluation over time using mutual gaze point prediction with joint visual attention (JVA) information. We then tested our proposed approach by extracting JVA features to evaluate team outcomes in a test user study. The results indicated that experimental teams interacted with 3-D digital learning tools had a high frequency of JVA and better knowledge retention than those in the control group. The teams identified by our automatic team assessment method to have higher JVA ratios were also found to perform better in knowledge tests after the collaborative intervention. We also investigated the association of user-study instrumental tools, and gender composition effects on JVA ratios and team test scores.

This work was a preliminary study of its kind to automatically assess team collaboration with computer vision techniques. Like any other project, there is room for improvement in the future. The focus of the current project was to understand collaboration on co-located situations, and shared gaze features were identified during the post-analysis process. There was a subset of collected data; since the main objective of this paper was to first understand dyadic interactions of students, so we have excluded larger teams. In future work, it will be ideal to evaluate the effectiveness of adopting our method to other collaborative learning scenarios. In further planned research, a dataset with team sizes larger than two is desired for better illustration of our idea and validation of our findings. Furthermore, we plan to work on multiple computer vision techniques by combining multiple image-based features, such as facial expression recognition (Li & Lam, 2015; Mollahosseini et al., 2016; Kar et al., 2019), emotion recognition (Tang et al., 2017; Motro et al., 2019), and head and body pose estimation (Ruiz et al., 2018; Pavlakos et al., 2019), along with joint attention estimation to more comprehensively interpret collaboration dynamics. The findings from this work offer implications in educational technology and collaborative computing fields by offering a novel assessment tool for team collaborations based on gaze information.

## Acknowledgments

We wish to express our sincere gratitude to the course instructors, research collaborators, laboratory technicians, teaching assistants, and study participants for their contributions to making this study possible.

## Statements on open data, ethics and conflict of interest

The study was approved by the Institutional Review Board (Protocol # XXXXXXXX). Since participant information was identifiable from the image data, we were not able to share participants' data from this study based on the ethical requirements. The authors declare no conflicts of interest.